\documentclass[12pt,preprint]{aastex}
\begin{document}

\title{Observations of Herbig Ae Disks with Nulling Interferometry$^{1,2}$}
\footnotetext[1]{The results presented here made use of the of MMT Observatory, 
a jointly operated facility of the University of Arizona and the Smithsonian Institution.}
\footnotetext[2]{This paper includes data gathered with the 6.5 meter Magellan Telescopes 
located at Las Campanas Observatory, Chile.}

\author{Wilson M. Liu$^{3}$, Philip M. Hinz, Michael R. Meyer, Eric E. Mamajek$^{4}$,
William F. Hoffmann, Guido Brusa, Doug Miller, and Matthew A. Kenworthy}
\affil{Steward Observatory, University of Arizona}
\affil{933 N. Cherry Ave., Tucson, AZ, USA 85721}
\email{wliu@as.arizona.edu}

\altaffiltext{3}{Michelson Graduate Fellow}
\altaffiltext{4}{Current address: Harvard-Smithsonian Center for Astrophysics, 
60 Garden St., MS-42, Cambridge, MA, USA 02138}

\begin{abstract}
We present the results of 10 $\mu$m nulling interferometric observations of 
13 Herbig Ae stars using the Magellan I (Baade) and the MMT 6.5 m telescopes.  
A portion of the
observations was completed with the adaptive secondary at the MMT. We have
conclusively spatially resolved 3 of the 13 stars, HD 100546, AB Aur,
and HD 179218, the latter two recently resolved using adaptive optics
in combination with nulling interferometry.  For the resolved objects
we find that the 10 $\mu$m emitting regions have a spatial extent of 15-30
AU in diameter.  We also have some evidence for resolved emission surrounding an
additional 2 stars (V892 Tau and R CrA).  For those objects in our study
with mid-IR SEDs classifications from Meeus et al. (2001), we find that the
Group I objects (those with constant to increasing mid-IR flux) are more
likely to be resolved, within our limited sample.  This trend is evident in
correlations in the inferred disk sizes vs. the sub-millimeter SED slope and
disk size vs. fractional infrared luminosity of the systems. We explore the 
spatial distribution and orientation of the warm dust in the resolved systems and
constrain physical models which are consistent with their observational 
signatures.
\end{abstract}

\section{Introduction}

Circumstellar disks have now been observed surrounding numerous 
pre-main-sequence (PMS) stars of intermediate mass.  Herbig Ae (HAE) stars are
of particular interest, as they are the evolutionary precursors to stars
such as Vega, which are known to harbor solid material in their circumstellar
environments.  Observations of HAE stars can therefore provide us with 
information regarding planet formation and the evolution of the circumstellar
disks as the stars evolve onto the main-sequence.  High spatial
resolution studies at wavelengths near 10$\mu$m are especially useful, as they can probe material
inner parts (few AU) of these disks, allowing one to directly observe
regions where planets could form in the habitable zones of such stars.

Infrared (IR) excess emission from HAE stars was originally explained by the
presence of geometrically thin, optically thick disks \citep{la92,hill92}, a
model that has been modified to include central star irradiation \citep{calvet},
disk flaring \citep{cg97,kh87}, 
and other structures to explain features in the spectral energy distributions (SED), 
such as an inner hole, inner wall heating, and self-
shadowing \citep{d02,ddn01}.  Alternative explanations for circumstellar 
emission take the form of dusty envelopes or envelope plus disk models 
\citep{m99,hkc93} as well as optically thin "haloes" in combination with disks
\citep{vink06,vink03}.  

A review of observational results prior to 2000 can be found in 
\citet{natta}.  More recently, a study by \citet[hereafter M01]{meeus} based on ISO
spectroscopy of HAE stars suggests that they may be classified into two groups
based on the shape of the SED, and explainable by differences in disk geometry.  We
compare our observational results to the M01 classifications in
\S \ref{sec-disc}.  Recent years have also seen the growth of interferometric 
observations of HAE stars (see \citet{m-g06} for a review).  
\citet{leinert} completed a long-baseline, spectrally
resolved survey of seven HAE stars using the Very Large Telescope Interferometer
at a wavelength of 10 $\mu$m.  Their study found that the 10 $\mu$m emission
regions to be 1-10 AU in size, and determined a correlation between the mid-IR 
SED slope and the physical size of the disk, where objects with larger emitting regions had redder SEDs.  
They have attributed this correlation to differences in disk geometry.  Additional
interferometric observations of HAE stars have been done in the near-IR and
include the studies of \citet{eis03} and \citet{m-g01}.  Eisner et al. used the Palomar
Testbed Interferometer at 2.2 $\mu$m and inferred the spatial structure and
orientation of the near-IR disks surrounding several stars.  In a study using
the Infrared Optical Telescope Array, Millan-Gabet et al. successfully resolved
several near-IR emitting regions surround HAE stars, establishing the a size
scale for the near-IR region of 0.5-5.9 AU, and finding that the distribution of
material surrounding resolved objects (including AB Aur) seem to favor a
circularly symmetric geometry.  We will compare our results to those of
previous interferometric studies in \S \ref{sec-disc}.  Observations at longer
wavelengths include a millimeter study of HAE stars by \citet{ms97} and
observations of AB Aur by \citet{marsh95}.

Nulling interferometry is a technique used to study spatially resolved 
circumstellar material in the presence of unresolved flux, and represents 
an ideal opportunity to observe the environments of HAE stars.  
The technique, first proposed by \citet{bracewell} is implemented by
overlapping the pupils of two telescopes (or two subapertures on a single
aperture telescope) with an optical path length difference
of one-half $\lambda$ between the beams.  The result of such a configuration is
a sinusoidal transmission function with the functional form
\begin{equation}
T(\theta) = sin^{2}(\pi b \theta/\lambda)
\end{equation}
where $b$ is the baseline of the interferometer, $\lambda$ is the wavelength
of observations and $\theta$ is the angular distance from the observed object to
the pointing center of the interferometer \citep{hinz01}.
During observations, the central destructive fringe is
placed on the unresolved point source.  This allows us to detect spatially resolved
emission, effectively isolating it from the unresolved stellar flux.  Thus,
nulling interferometry provides the necessary contrast to observe faint
circumstellar material in the presence of a much brighter star.  In
observing pre-main sequence stars at wavelengths near 10 $\mu$m, where the 
circumstellar emission dominates the stellar flux (often by more than 2 orders
of magnitude), isolating the resolved disk emission from the unresolved disk 
emission gives us valuable information about the spatial
distribution of circumstellar dust.  This technique can detect material as 
close to the star as one-quarter of the fringe spacing where the light is neither suppressed nor enhanced.  This corresponds to 0.12 arcseconds for the configuration used on the MMT, or 12 AU about a star at 100 pc.  This is between 2 and 3 times finer spatial resolution than the diffraction limit
at the wavelengths observed in this study (8 to 13 $\mu$m).

In more recent observations at the MMT, nulling interferometry was implemented in combination with the MMT's adaptive optics (AO) secondary mirror.  The addition of AO has benefits for nulling.   Specifically, wavefront abberrations introduced by the atmosphere that might affect the suppression level of the null are significantly reduced or eliminated.  The stabilization of the incoming
wavefront therefore allows us to precisely tune the path length between the arms of the
interferometer, allowing for the best possible suppression of light, hence
observations are made more efficient (see \S \ref{sec-obs}).  Additionally,
a stable wavefront and image allows us to integrate for greater
periods of time, making fainter resolved flux detectable.  The unique nature of the deformable secondary also has benefits in mid-infrared
observations.  Since the secondary mirror of the telescope is the deformable 
mirror, there is no need for an intermediate set of correcting optics between
the telescope and science camera.  This reduces the number of warm surfaces in
the optical path, minimizing background and maximizing throughput.  More technical
details regarding the MMT adaptive secondary can be found in \citet{brusa}.

In \S \ref{sec-obs} we discuss the target sample and observations, as well
as the reduction procedure for nulling interferometric data.
In \S \ref{sec-res} we present the results of the observations and discuss
them in context of models and previous observations in \S \ref{sec-disc}.

\section{Observations and Data Reduction} \label{sec-obs}
\subsection{Target Selection}
Targets for this survey were taken from a catalog of Herbig Ae/Be stars by
\citet{the94}.  The objects were chosen to include stars with a spectral
type of B8 or later, as HAe stars have been found to have a significantly higher 
incidence of circumstellar disks than Herbig Be stars \citep{natta}.  
All objects (except one, HD 98922) were chosen to be within 250 pc, to ensure 
our ability to spatially resolve a circumstellar disk, should one be
present.  All targets also have a 12 $\mu$m flux (IRAS) greater than 10 Jy, to
insure sensitivity to any resolved emission.  A large flux is necessary for non-AO
observations, as we are limited to short integration times (see \S \ref{sec-data}).
The final sample of 13 stars, along with their stellar characteristics, is shown
in Table \ref{tab-sample}.

Observations of 13 Herbig Ae stars were made in 2001 August and 2002 May at the 
6.5 m Magellan I (Baade) telescope at the Las Campanas Observatory, Chile and
2002 November, 2004 January, and 2005 June at the 6.5 m MMT at Mt. Hopkins, Arizona.
On these single-aperture telescopes, nulling interferometry is implemented by
dividing the aperture into two identical elliptical subapertures, each
2.5 x 5 m with a baseline of 4 m.  The BracewelL Infrared Nulling Cryostat
(BLINC; \citet{hinz_thesis}) is the interferometer which re-images the primary mirror 
as two subapertures, and recombines the beams which an appropriate path difference.
The recombined image is fed to the Mid-Infrared Array Camera 3 (MIRAC3;
\citet{hoffmann}),
which provides the final stop for the image.
Observations were taken at 10.3, 11.7, and in a few cases 12.5 $\mu$m, each with a 
bandwidth of 10\%, 
and a wide 10.6 $\mu$m (N-band) filter with a 50\% bandpass.   
Additionally, observations of HD 100546
taken at Magellan I included direct imaging taken with BLINC-MIRAC at longer
wavelengths (18.0 and 24.5 $\mu$m).

\subsection{Non-AO Observations} \label{sec-non-ao}
Observations taken at Magellan and during the 2002 MMT run were done without AO.
Without wavefront correction, atmospheric aberrations randomize the path difference
between the two arms of the interferometer.  Thus data must be taken
with short frame times (50 ms) in order to freeze out the seeing effects.  In
this case, images are taken contiguously in large sets, typically 500 to a set.
This number of frames is sufficient to sample the phase between the beams, and thus include several frames with a phase difference close enought to one-half wave that higher order spatial errors dominate the level of residual flux in the focal plane.
The images with the best null (destructive interference) and constructive 
interference in each set are selected in order to evaluate the "instrumental null" 
(see below).  Ten off-source sky frames, used to subtract out the
sky background, are taken after each
set of frames on-source.  For each science object we take several sets
of frames, in order to effectively evaluate the null.  From the lowest and
highest fluxes in each set of frames, we calculate the \emph{instrumental null}
which is defined as:
\begin{equation}
N = Flux_{nulled} / Flux_{constructive}
\end{equation}
and expressed as a percent.

Observations of each
science object are followed by 2 to 4 sets of 500 frames of a point source
(spatially unresolved) calibrator star in order to establish a baseline for goodness of null achieved
in the most recent science observations.  Calibrator nulls generally ranged between
5\% and 15\%. We use the null achieved on the
calibrator to calculate the \emph{source null} defined as the difference
between the instrumental null on the science object and the calibrator:
\begin{equation}
S = N_{science} - N_{calibrator}
\end{equation}
This represents the resolved flux as a percentage of the
full flux of the target.  A source null of zero means that the science object
is spatially unresolved. 

Additionally, observations are taken with different rotations of the
interferometer baseline relative to the sky.  Typically, two sets of 500
frames are taken at each rotation. This allows us to probe for the
presence of an elongated or flattened structure (such as an inclined disk) if
resolved emission is detected, which is accomplished in the following manner.  The
output of the interferometer is a transmission pattern superimposed on the
stellar image.  This transmission pattern has a sinusoidal functional form shown in \S 1, with the
destructive interference fringe directly on axis, and aligned perpendicular
to the baseline of the interferometer.  If a flattened extended structure is
present surrounding our science target, and the major axis of the emission
is aligned parallel to the destructive interference fringe, we would expect
a deeper null (smaller $N$) than when the fringe and the major axis are
orthogonal.  The resulting dependence of $N$ vs. the rotation of the
interferometer would be sinusoidal with a period of 180$\degr$,
\begin{equation}
S = a + b * sin(PA + \theta)
\end{equation}
The parameters $a$, $b$, PA, and $\theta$ are determined by the size of the
emitting region,
its inclination, the position angle of its major axis, and rotation of the
interferometer baseline, respectively.

\subsection{Observations with AO}
Observations taken in 2004 January and 2005 June at the MMT were done with
the AO secondary.
As described in the Introduction, the use of AO significantly increases the
efficiency of observations.  Since, with AO, we are able to precisely tune the
path difference between the arms of the interferometer, large sets of fast
frames to freeze-out atmospheric effects are unnecessary. AO also stabilizes
the image, generally resulting in smaller errors in the null. 
In this case, we take sets of 10 frames (usually $\sim 1$ s integration)
of the object tuned to destructive interference, followed by 10 frames of the
object in constructive interference.  Observations of the calibrator are taken
in the same manner, with off-source sky frames taken after each 
destructive-constructive pair.  Calibrator nulls typically ranged between 3\% and
6\%, indicating an improvment in both null depth and stability compared to
non-AO observations.

Table \ref{tab-obs} summarizes all observations of science targets, taken 
with and without AO.

\subsection{Nulling Data Reduction} \label{sec-data}
For non-AO observations, each on-source frame is sky subtracted using a sky frame
created by median combining the 10 off-source frames taken immediately after the 
science frames.  A custom IDL program is used to extract aperture photometry 
from each on sky subtracted frame.  The aperture size for each set of frames is
chosen by determining the radius at which the source flux disappears into the
noise.  The noise level is assessed in an annular region several pixels wide
well outside the aperture.  For each set of 500 frames, we identify the frames 
with the smallest residual flux (best null) and the brightest constructive image.  
These two frames are used to evaluate the instrumental null.  The null for
each set of calibrator images are evaluated in an identical way.  

For the frames taken with AO, the best destructive frame in each set is determined
using aperture photometry in a manner identical to the non-AO procedure described
above.  The instrumental null is determined by using the best nulled image
and a median combined composite image of the constructive
frames.  Instrumental nulls for calibrator stars, taken in between science
objects, were evaluated in an identical manner.

Source nulls were evaluated for all data by subtracting the instrumental null
of the calibrator from the instrumental null achieved on the science object.
A table of instrumental, calibrator, source nulls, and errors is shown in
Table \ref{tab-nulls}.  The values of the nulls represent an average of the
data sets taken at each wavelength and position angle, and errors are the
1 $\sigma$ dispersion in the values.

\section{Results}\label{sec-res}
Out of 13 objects observed, we have conclusively resolved three of the objects,
HD 100546 (at 3 different wavelengths and 5 different rotations) and AB Aur
(at one wavelength and 5 different rotations) and HD 179218 (at one wavelength
and 2 rotations).  Additionally, we have marginally resolved (at about the 2.5
$\sigma$ level) two stars, V892 Tau and R CrA.  Source nulls for all objects can be 
found in Table \ref{tab-nulls}.

One may note that source nulls are negative at a significant level
($>2\sigma$)  for one set of data.  Negative source nulls are unphysical,
hence the appearance here necessitates an explanation.  The negative value
for the set in question (HD 98922, 11.7 $\mu$m) can be attributed
to an inaccurate measurement of the calibrator null taken immediately
after the science object.  Analysis of the calibrator images shows
a slightly "dual-peaked" image of the star, indicating a slight misalignment
in the beams of the interferometer and/or a temporary degradation in
observing conditions (i.e., bad seeing).  This results in a poor null
(abnormally large value) for the calibrator null, hence a negative value
for a source null.  It is important to note that a poor null in the
calibrator cannot result in a false positive detection of resolved material,
and any detections of resolved emission are checked to ensure that
they are not a result of poor seeing or misalignment.  The negative source
nulls are included here for the sake of completeness in presenting data.
Other negative source nulls are present in the results, but their significance
is marginal, as the values are negative at less than a 2 $\sigma$ significance.
The large values of calibrator nulls and errors in these cases are likely
due to short-lived poor seeing affecting a single observation of the
calibrator (two or more observations of the same calibrator are averaged to
produce the quoted value for the calibrator null).  Again, the inclusion of
these results is for the sake of completeness and transparency in our
methodology.

\subsection{Simple Disk Models - Assessing Size and Flaring \label{sec-disc2}}
In order to infer the size of the emitting region for 
each resolved object, we use two simple disk models: 1) the intensity of the 
disk is a Gaussian function with the peak at the center; and 2) the source of the
emission is confined to an annulus  of uniform intensity around the star.  
The models are placed at the distance of the object and convolved with the
transmission pattern of the interferometer, and a theoretical source null is
calculated.  The size of the emitting region in the model is adjusted until the 
results best match the observationally determined source null.  Models are fit
for each wavelength at which an object was observed.  For any object which
we have data at several rotations of the interferometer baseline, we fit a
sinusoidal function of the form described in \S \ref{sec-non-ao} to the $S$ vs.
PA relation determined from our nulling observations.  The fitting
procedure is described in detail in \citet{liu03} and \citet{liu05}.

The extent of the 10 $\mu$m emitting region in each of the resolved objects
is a powerful tool for determining the physical structure of warm dust in the
system.  Several factors contribute to the spatial extent of emission. 
One factor is the degree of disk flaring, thus we explore the effect of flaring
on the nulling observations.  Another factor affecting the extent of the emitting
region is grain size, as larger blackbody grains would tend to result in
an emitting region closer to the star, due to the relative efficiency in emission compared to small grains.  We also examine the effect of grain size on our observations.

\begin{figure}
\epsscale{0.85}
\plotone{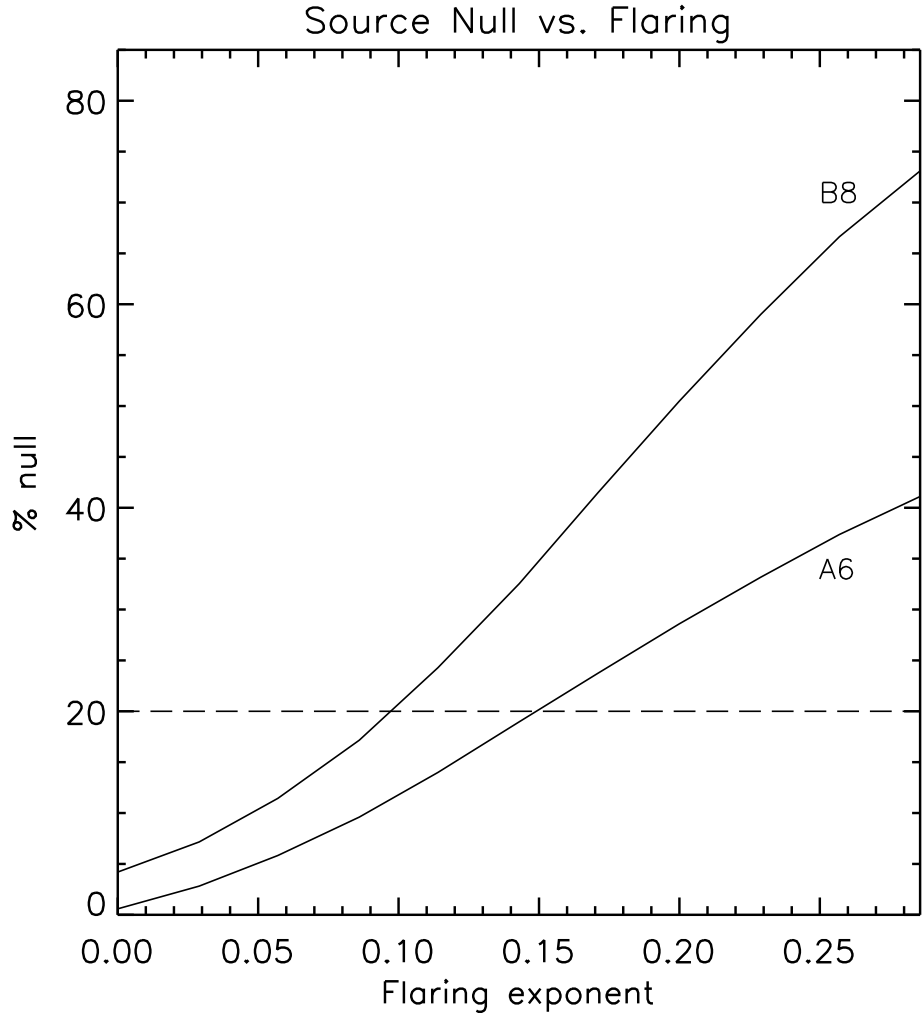}
\caption{Predicted source null vs. degree of flaring.  The degree of flaring is
represented by varying the exponent of $a$ in \citet[Eq.10]{cg97} from 0 
(the flat disk case) to 2/7 (the vertical hydrostatic equilibrium case) 
Spectral types A6 and B8 are shown; intermediate types lie between the two curves.
The horizontal dashed line represents the median (and mean) source null (20\%) for
all spatially resolved measurements.}
\label{fig-flarenull}
\end{figure}

In order to assess the effect that the degree of disk flaring has on the
resolvability of a target, we construct a simple face-on flared disk model based
on the model of \citet[hereafter CG97]{cg97}.  In our model we vary the degree of flaring by adjusting the power-law dependence of the flaring term, $H/a$ 
(Eq. 10, CG97, where $H$ is the height of the disk above the midplane and
$a$ is the radial separation between the star and a point on the midplane of the disk). The exponent of $a$ is varied from 0 (the flat disk case, i.e. constant H/a)
to 2/7 (the vertical hydrostatic equilibrium case, i.e. $H/a \sim a^{2/7}$).  
Other parameters assumed in the CG97 model include a dust mass of 1\% of the gas
mass, uniformly mixed, a grain size of 0.1 $\mu$m, and a mass density of 2 g cm$^{-3}$.
The CG97 model was used to determine the brightness from a hypothetical disk,
using canonical stellar parameters (effective temperature, mass, radius and 
luminosity) \citep{cox} appropriate for the spectral types of the targets' 
parent stars, while varying the amount of flaring in the disk.  This brightness signature is
then convolved with the transmission function of the interferometer to
predict the source nulls expected for varying degrees of flaring.  Figure
\ref{fig-flarenull} shows the predicted values of the source null as a
function of the degree of flaring (represented by the value of the exponent
of $a$) for a range of spectral types.  We see that the model predicts very 
large source nulls (40-75\%)
for the maximally flared disks in all cases.  This holds true also for the
outer part of the model \citep[Eq.5]{ddn01} which predicts flaring
as the disk emerges from the shadowed region. The largest source
null for our resolved objects approaches 40\% for HD 100546, a B9 star.
Thus, it appears that if one considers flaring as the only factor
affecting the observed source null, none of the objects display flaring
to the degree of the hydrostatic equilibrium case.  It is a possibility that
this is an indication that the warm dust may be in a flatter distribution
than expected (possibly due to dust settling toward the disk midplane). However, 
one must keep in mind that other factors (such as the orientation of the
dust relative to the rotation of the interferometer baseline) can have a significant effect on the source null. 

Spatial information about the emission regions surrounding these stars is
also important in breaking degeneracies in interpreting the SEDs with regard
to grain size. For example, large grains, with sizes similar to the wavelength of
emission, at closer separations the the star can manifest themselves in the
same way as small, ISM sized grains in a more extended distribution.  In the 
model based on CG97 described above, the assumed grain size is 0.1 $\mu$m, or
ISM-sized grains.  Thus, for resolved objects, if grain sizes are actually 
larger than the assumed size, this would require a greater degree of flaring 
than the actual values calculated with the model, in order to account for the 
same level of resolved flux.  When comparing the relative effect that grain
size and flaring has on the temperature profile of the dust, we
see that varying the flaring has a greater effect on the power law dependence
of the effective temperature vs. separation from the star.  Increasing the flaring from a flat distribution
to a flared distribution (at vertical hydrostatic equilibrium) results in a
change from a $T \sim r^{-0.75}$ relation to $T \sim r^{-0.4}$, and
to a change in source null of 40-70\%, depending on the luminosity of the star. 

Table \ref{tab-models} summarizes the sizes and flaring parameters that best
fit the nulling data, with the errors representing the error in source null
shown in Table \ref{tab-nulls}. For unresolved objects, a maximum size for the
emitting region is shown, calculated using the error in the source null.

\subsection{Notes on Resolved Objects}\label{sec-notes}

\textbf{\emph{HD 100546}} \
This object is perhaps the oldest star in our sample, with an estimated age exceeding 10 Myr (see footnote, Table \ref{tab-sample}).  Large fractions
of crystalline silicates suggests an evolved disk, and evidence of
a giant protoplanet in the system \citep{bouwman,malfait,wael}.
HD 100546 was resolved with non-AO observations (both nulling interferometric
and direct imaging) at Magellan I.  Nulling observations at 10.3, 11.7, and
12.5 $\mu$m show evidence for an inclined disk with an orientation of
$45\degr$ from face-on and a radius of about 12 AU, with a slightly larger
size resulting from the ring model \citep{liu03}.  Direct imaging at
18.0 and 24.5 $\mu$m also show resolved emission at a greater separation
(15-20 AU) from the star, and verify the orientation of the dust disk.
We find that the relative sizes of the 10 and 20 $\mu$m emission to not
agree with the $T \sim r^{-0.5}$ relation for a continuous flared disk
(CG97).  Instead, a disk with a large inner clearing($r < 10$ AU)
would result in the 10 $\mu$m emission being detected at a greater
separation than expected, explaining the relative sizes of the emission
at different wavelengths.  This inner clearing could be the result of
the formation of a giant planet. Further discussion of this
model can be found in \citet{liu03}.  Here we also consider the possibility 
that the relative sizes of the emission regions may be a consequence of
an inner self-shadowed region in the disk.  In the models of \citet{ddn01}, 
one sees that for the shadowing to extend out to the radii of the detected 
10 $\mu$m emission ($>$10 AU), the height of the inner rim would have to be 
enhanced relative to the height expected, due to direct normal incidence
radiation.
If this was the case, the models also predict a significant decrease in the
strength of the 10$\mu$m emission feature.  As HD 100546 shows a strong
emission feature, it appears that self-shadowing is not a
significant effect in determining the relative sizes of the the resolved
emission regions in this system.  The relatively large source nulls
of HD 100546 suggestive a moderate degree of flaring even with the smallest
grain sizes.  This result supports the fact that self-shadowing is not a factor
in the regions we observe and is also in agreement with the SED analysis
of \citet{bouwman}, which infers that the vertical height of the disk
must be enhanced at $\sim 10$ AU.

\textbf{\emph{AB Aur}}\ \label{sec-abaur}
The age of AB Aur is estimated to be significantly younger than HD 100546,
with its SED showing no evidence for crystalline silicates (M01).
Observations of AB Aur at 10.3 $\mu$m (N-band) with AO were taken at 5 
different rotations of the interferometer baseline and, like HD 100546, 
show a variation
consistent with the presence of an inclined disk.  A fit to the data
results in a disk radius of 12-15 AU, with the smaller sizes in this range from
the Gaussian disk model, and the larger sizes a result of fitting to the
ring model. These sizes are consistent with a recent mid-IR study by 
\citet{marinas}.  A disk inclination of $45\degr$ to
$65\degr$ and a major axis PA of $15\degr$ to $45\degr$ are also
inferred \citep{liu05}.  The orientation
of the disk (inclination, PA) derived from these mid-IR observations differ
from several near-IR and millimeter observations previously completed
\citep{f04,eis03,m-g01,ms97}.  This suggests that the circumstellar
environment may be more complex than a simple disk structure.  However,
the sizes of the disk at different wavelengths from this and
the aforementioned studies, in addition to the mid-IR study of
\citet{cj03}, indicate that the wavelength vs. size relation is consistent
with a temperature profile of $T \sim r^{-0.5}$ expected from a flared
disk.  A full discussion can be found in \citet{liu05}.  From relatively
large source nulls and the fit to the flaring model, a moderate amount of 
flaring can be inferred.

\begin{figure}
\epsscale{0.85}
\plotone{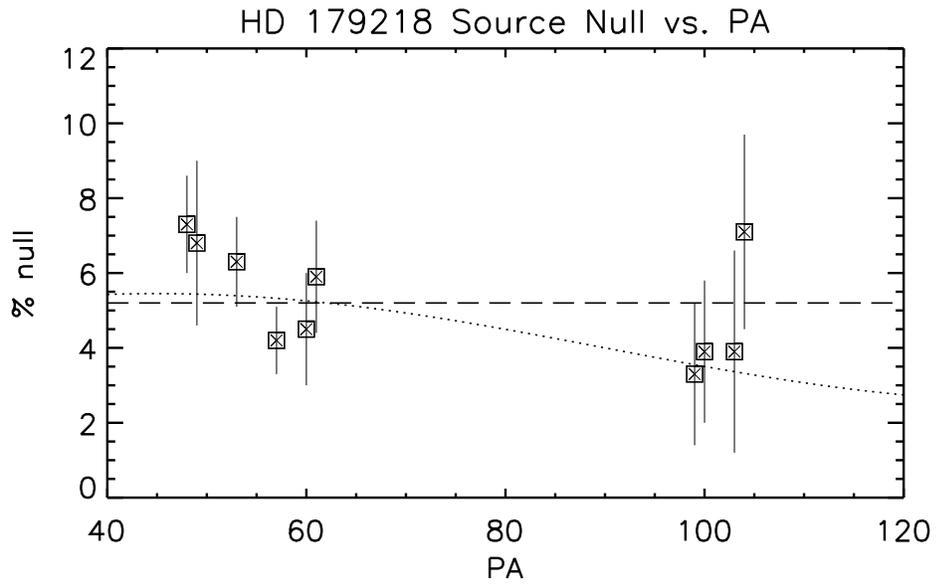}
\caption{Source null vs. PA for HD 179218, spatially resolved with AO observations.
The horizontal dashed line is the mean of the measurements, and expected signature
(no variation in null vs. PA) from a circularly symmetric source.  The dotted line
shows the expected variation in null for a 20 AU diameter, Gaussian intensity
distribution inclined at $45\degr$.}

\label{fig-hd179218}
\end{figure}

\textbf{\emph{HD 179218}} \
One of the most distant object in our sample (244 pc; \citet{perryman}),
the SED of this object shows significant levels of crystalline silicates
(M01).
Initial observations without AO failed to resolve this star, resulting
in a source null of $3 \pm 3\%$ at a PA of $162\degr$ \citep{hinz01}.  Given the greater
precision with the use of AO, follow-up observations were made in 2005 June.
These follow-up observations have positively detected resolved emission at
a levels from 3 to 7\% (0.7 to 1.6 Jy).  A plot of source null vs. PA is
shown in Fig. \ref{fig-hd179218}.  An average of the source nulls 
implies a FWHM of $20 \pm 4$ AU for the Gaussian disk model or diameter
of $27 \pm 5$ AU for a ring distribution of dust.  
The low source nulls may also suggest flaring in the
dust which is small; significantly less than the vertical hydrostatic 
equilibrium case.  
The $S$ vs. rotation relation does not show significant variation, consistent with
circular symmetry, though significant inclinations cannot be ruled out, as a 
hypothetical disk with an inclination of $45\degr$ results in a
variation in the source null of about 3\%, within the errors of the measurement.  
The presence of
significant silicate emission in the ISO spectra of M01 also seems
to rule out a large inclination for the object, which would result in a drop off in emission intensity at wavelengths shorter than about 10 $\mu$m \citep{cg99}. 
However, we do not make any definite conclusions regarding the spatial 
orientation of the dust.

\textbf{\emph{V892 Tau}} \
Observations at 11.7$\mu$m without AO, and at 10.3$\mu$m, using AO, show 
resolved emission at a level of about 3 Jy
from this Herbig Ae source.  The emission is detected at a PA of $164\degr$,
but no information can be derived as to the overall orientation of the
emission, as data was taken at only one rotation.  Using an average of
the source nulls obtained in the two data sets at 11.7 $\mu$m and assuming a 
Gaussian intensity distribution for the dust, the FWHM is 18-28 AU.  The ring
distribution yields a diameter of 25-37 AU.  At 10.3 $\mu$m, these sizes are
12-16 and  17-23 AU, respectively.  The relative sizes of the emission at the two
wavelengths is consistent within errors to the expected $T \sim r^{-0.5}$ relation
for a CG97 flared disk, assuming purely thermal emission.  
Signs of flaring can also be inferred from the higher level of resolved emission 
in the 11.7 $\mu$m observations, but, when considering flaring as the sole factor
affecting the extent of the emission region, suggests that the dust lies in a 
flatter distribution than the vertical hydrostatic equilibrium case, though this
may also be an indication of larger grains.  

\textbf{\emph{R CrA}} \
This object was observed without AO at Magellan I and shows
marginal evidence for resolved emission.  The object
was observed at two rotations of the interferometer and one
of the two rotations yielded a positive detection at the 2$\sigma$ level.
The level of resolved emission is 8\% and suggests a
spatial extent for the dust of $15 \pm 4$ AU using a Gaussian
dust distribution and $20 \pm 4$ AU using a ring distribution.  
As with V892 Tau, the source nulls are possibly indicative of less 
flaring than a hydrostatic equilibrium situation would expect, or larger
grains than ISM sizes.

\section{Discussion}\label{sec-disc}

\subsection{Trends in Resolved Objects}
\begin{figure}
\epsscale{0.85}
\plotone{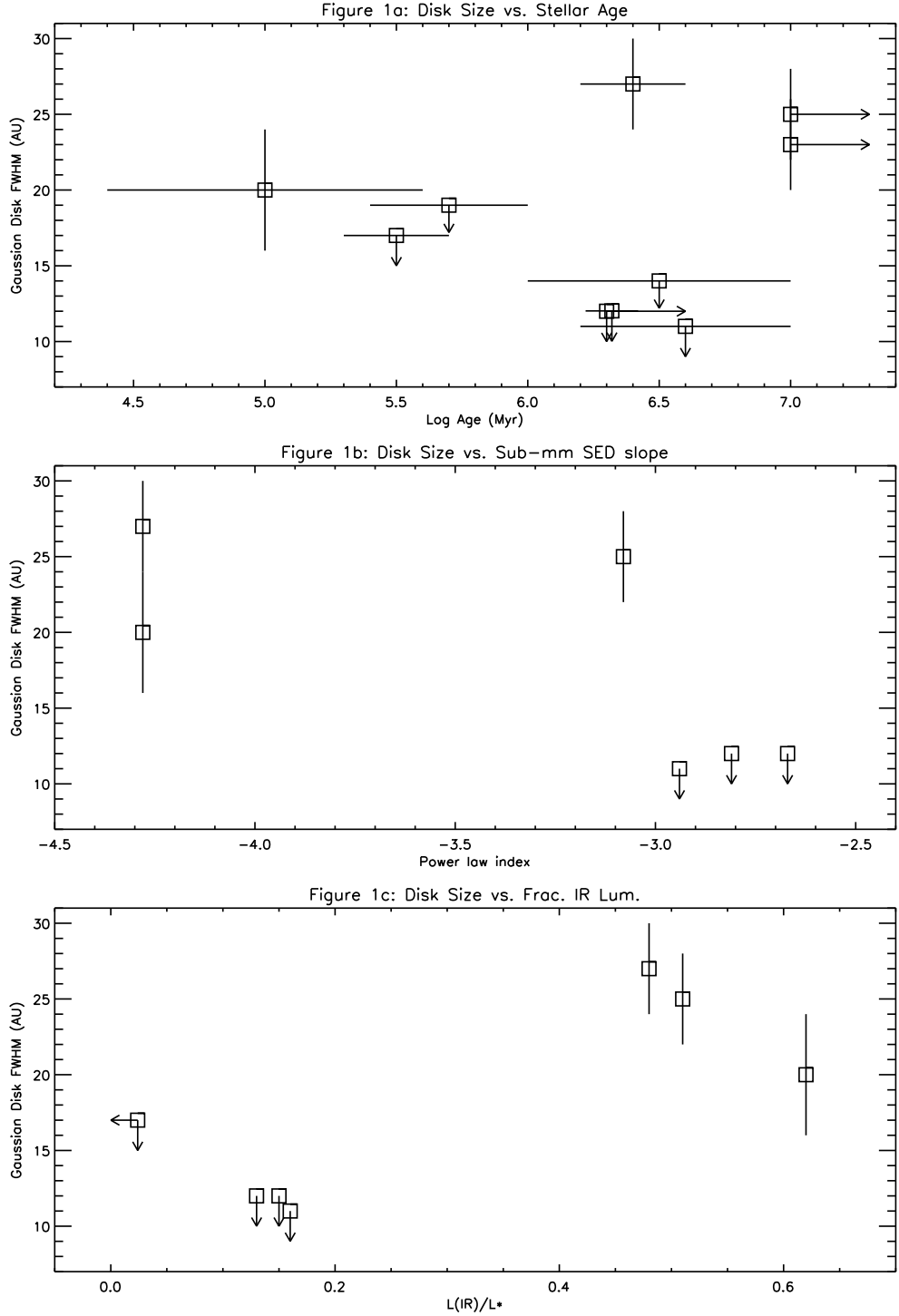}
\caption{\textbf{(a)} Disk size (inferred from fitting a Gaussian intensity distribution to the source null) vs.
stellar age (for ten objects with age determinations in the literature) for
our sample.  Sizes plotted are derived from the 11.7 $\mu$m observations except
AB Aur (10.3 $\mu$m) and HD 179218 (10.6 $\mu$m).  No clear age trend is apparent
in the data.
\textbf{(b)} Disk size vs. sub-mm SED slope (for the six objects with values determined in
M01).  Objects with steeper SEDs in the sub-mm (index $<$ -3.0) correlate with Group I objects and appear to have larger, resolvable disks.
\textbf{(c)} Disk size vs. fractional IR luminosity (seven objects from M01).  A larger
IR excess correlates to larger disk size.  Given the poor constraint on the distance
to HD 98922, it is not included in any of the plots.}
\label{fig-agesize}
\end{figure}

In selecting our HAE targets, we included objects with a range of spectral types
and ages, in the hope that any evidence of evolution in the PMS environment, or 
differences in the circumstellar region due to the stellar mass
of the parent star could be probed.  
However, we have found that the resolved objects in our sample 
appear to have a range of stellar characteristics.  The resolved objects
have spectral types ranging from late-B type (HD 100546 and HD 179218) to A6 (V892 Tau), 
and ages ranging from approximately 0.1 (HD 179218) to $>10$ Myr (HD 100546).  
Therefore, there seems to be no distinguishing characteristic between unresolved 
and resolved sources in terms of age or spectral type.  A plot of the inferred disk
size (see \S \ref{sec-disc2}) vs. the stellar age 
(for those objects with age determinations in the literature)
is shown in Figure \ref{fig-agesize}a and shows no obvious trend.  This could be due
to either: 1) errors in age estimates  for these stars or 2) the fact that time is not the sole or dominant factor affecting disk evolution.

An analysis of the spectral energy distributions (SED) of the observed
stars, however, does suggest a trend in the resolved objects.  M01 categorizes
Herbig stars into two major groups: Group I with large amounts of mid-infrared
excess, and Group II with moderate quantities of mid-IR excess, descending
at longer wavelengths.  For those objects with classifications in the literature, 
we find that two out of three Group I objects were resolved in 
our initial observations (we do not include HD 179218 as a resolved object here
since it was unresolved in our initial, non-AO observations), 
while zero of five Group II objects showed resolved emission.  
This trend is evident in a plot of disk size vs. the sub-millimeter
SED slope (Fig. \ref{fig-agesize}b), which M01 finds correlates well with the mid-IR SED grouping of the star and could be considered a surrogate for the evolutionary state of the disk.  Objects with steeper sub-mm slopes (index $<$ -3) appear to have
larger, resolvable disks.  A calculation of the Kendall's $\tau$ correlation coefficient \citep{num_rec} for the SED slope vs. disk size relation yields a correlation probability of 94\%, whereas the stellar age vs. disk size relation discussed in the previous
paragraph yields a correlation probability of 43\%.  We also see a correlation (87\%
probability) between the fractional IR luminosity (calculated by M01) and the disk 
size (Fig. \ref{fig-agesize}c).  

M01 attributes the difference in the SED
between these two groups to disk geometry, with Group I objects displaying
a significant amount of flaring outside of the inner disk, while disks
in a Group II source have less flaring, a result of shielding from an
optically thick inner disk, or perhaps self-shadowing from a puffed up
inner wall \citep{ddn01}.  It is conceivable that the flaring in
the Group I objects results in the disk intercepting more radiation at
greater radii, making it easier to spatially resolve the dust disk.
M01 correlates the amount of mid-IR excess to the scale height
of the disk, with greater excesses a result of more substantial flaring
in the disk, consistent with the observed trend in our survey.
In contrast, the 10 $\mu$m emission in the Group II shadowed disk will
be confined to the inner few AU, making it more difficult for resolved
emission to be detected.  Thus our results are in good agreement with those
of \citet{leinert} which found that the mid-IR emitting regions were
larger for the redder, Group I objects.

Despite this line of reasoning, there is reason
to be cautious before attributing the characteristics of the SED groupings
to an all-encompassing physical model.  M01 make the assertion
that Group II objects show evidence that they are more evolved, due to
large grain sizes.  This would seem to imply that older objects are
less likely to be resolved than younger ones, if time is the dominant
factor in the evolution of these systems.  In our sample, we do
not find any trends in resolved objects with age.  This would imply
that either the age of the system is a poor indicator of the evolutionary
state of the system, or that the ages attributed to the stars are in error.
We can gain insight into this issue by comparing HD 100546 and
AB Aur, the two most conclusively resolved objects.  By comparing
observations at different wavelengths for each object from this and
other studies, we find that the temperature vs. radius relation for the
stars is dramatically different.
The temperature profile suggests that AB Aur and HD 100546 do not
have similar circumstellar environments despite the similarity in the
10 $\mu$m resolved emission.  The former is consistent with the
$T \sim r^{-0.5}$ relation expected from a simple flared disk CG97
while the latter shows evidence for a inner disk gap 
(see details in \S \ref{sec-notes} and
\citet{liu03}).  Thus, the evidence here suggests that our ability
to resolve an object may be due to different circumstances in each system
and cannot necessarily be attributed to similar physical models.

\section{Summary and Conclusions}
We have carried out 10 $\mu$m nulling interferometric observations of 
13 Herbig Ae stars, and reach the following primary conclusions:

- We have conclusively resolved warm dust surrounding 3 objects, HD 100546,
AB Aur, and HD 179218, the latter previously unreported.  Both HD 100546 and
AB Aur show significant variation in source null vs. PA, which is evidence for
an elongated structure such as an inclined disk \citep{liu03,liu05}.  HD 179218
was resolved in recent (2005 June) observations with AO, and preliminary results
suggest little variation of null vs. PA, consistent with an axisymmetric
distribution (such as a face-on disk) for the dust, though a significant inclination cannot be ruled out given the errors in the measurement.

- We have found evidence for resolved emission around an additional two HAE stars, V892 Tau and R CrA.  Both the sources show resolved emission at a level of a few percent of the unresolved flux at 10 $\mu$m.

- The spatial extent of the emitting region in the resolved systems range from 
15-30 AU in diameter, assuming two models: a Gaussian disk and a ring.

- Both SED slope and fractional IR luminosity appear to
be good indicators of the spatial extent of circumstellar dust.
Although our sample size is small, it appears that M01 Group I objects 
are more resolvable than Group II objects, a result consistent with \citet{leinert}. This trend is evident in the correlation between disk size vs. sub-mm spectral slope as well as disk size vs. fractional IR luminosity.

- There is a lack of correlation in disk size vs. stellar age, perhaps due to
uncertainty in age determination, and/or the fact that time is not the sole or dominating factor in disk evolution.  

- Using a model based on CG97, we evaluate the effect of disk flaring
on the resolvability of the objects.  We find that a hypothetical object
with flaring consistent with vertical hydrostatic equilibrium would produce
a very large source null in all objects.  The fact that the source nulls observed in these objects is not that large suggests that flaring does not impact the resolvability of the objects as much as expected.  In this latter case, an alternative explanation for the correlation between the SED groups and resolvability must be determined. One possible explanation for this is the grain size.  With grain sizes larger than ISM sizes, flaring must be enhanced to account for the same amount of resolved flux.

- Follow up AO observations should be made on all objects not yet observed with
AO, in order to better constrain the limits, or possibly detect, their spatial
extent.  Observations of Group Ib objects, those without strong 10 $\mu$m silicate emission, would also be interesting to assess any correlation between the presence of these emission lines and the spatial resolvability of the objects, and further constrain the distribution and orientation of the disks.

\section{Acknowledgments}
W.M.L. was supported under a Michelson Graduate Fellowship.  E.E.M. is supported
through a Clay Postdoctoral Fellowship from the Smithsonian Astrophysical Observatory.  We are grateful to the
operators and staffs at the MMT and Magellan Observatories for their support of our observations.  We thank A. Breuninger and B. Duffy for technical support of
BLINC-MIRAC.  The authors also thank J. Najita and the anonymous referee for helpful comments in the revision of this paper. P.H. and M.R.M. acknowledge support from the NASA Astrobiology Institute.  BLINC was developed under a grant from NASA/JPL.  The MMT AO system was developed with support from the Air Force Office of Scientific Research. This work made use of the SIMBAD database.

\clearpage
\begin{deluxetable}{cccccc}
\tablecaption{Target List \label{tab-sample}}
\tablewidth{0pt}
\tablehead{
\colhead{Name} &
\colhead{Spec. Type} &
\colhead{d(pc)} &
\colhead{log Age(Myr)} &
\colhead{Group$^{*}$} &
\colhead{Refs.}
}
\startdata
HR 5999 & A5-7 & $210 \pm 40$ & $5.7 \pm 0.3$ & & 1,3,4,6\\
KK Oph & A6 & $160 \pm 30$ & $6.5 \pm 0.5$ & II & 2\\
DK Cha & A & $\sim200$ & & & 1,3 \\
HD 150193 & A1 & $150 \pm 30$ & $>6.3$ & II & 1,4,6\\
HD 98922 & B9 & $>540$ & & & 1,4\\
HD 104237 & A4 & $116 \pm 8$ & $6.3 \pm 0.1$ & II & 1,4,6\\
51 Oph & A0 & $131 \pm 15$ & $5.5 \pm 0.2$ & II & 1,2,6\\
R CrA & B8 & $\sim130$ & & & 3\\
AB Aur & A0 & $144 \pm 20$ & $6.4 \pm 0.2$ & I & 3,4,5,6\\
V892 Tau & A6 & $150 \pm 10$ & $>7$ & & 3,5\\
HD 100546 & B9 & $103 \pm 7$ & $>7^{\dag}$ & I & 1,2,4,6\\
HD 163296 & A3 & $122 \pm 15$ & $6.6 \pm 0.4$ & II & 2,6\\
HD 179218 & B9 & $244 \pm 55$ & $5.0 \pm 0.6$ & I & 2,6
\enddata
\tablecomments{References- 1) SIMBAD; 2) \citet{leinert}; 3) \citet{hama}; 4) \citet{vdA}; 5) \citet[and ref. therein]{natta}, 6) \citet{perryman}.\\
$^{*}$\citet{meeus} SED Group\\
$^{\dag}$HD 100546 attributed to Lower Centaurus-Crux
OB association by \citet{dezeeuw} with an association age of 16 Myr 
\citet{mamajek}.
}
\end{deluxetable}

\clearpage
\begin{deluxetable}{cccccccc}
\rotate
\tablecaption{Summary of Observations \label{tab-obs}}
\tablewidth{0pt}
\tablehead{
\colhead{Date} &
\colhead{Star} &
\colhead{$\lambda$ ($\mu$m)} &
\colhead{\# Frames} &
\colhead{Intgr./frame (s)} &
\colhead{PA (or rotation$^{*}$)(\degr)} &
\colhead{Tel.} &
\colhead{Ref.}
}
\startdata
1-2 & HR 5999 & 11.7 & 1000 & 0.5 & -165(rot.) & Mag. & 1\\
3-4 & " & " & " & " & 110(rot.) & " & 1\\
5-6 & KK Oph & 11.7 & 1000 & 0.5 & -150(rot.) & Mag. & 1\\
7-8 & " & " & " & " & 120(rot.) & " & 1\\
9-10 & DK Cha & 11.7 & 1000 & 0.45 & -170(rot.) & Mag. & 1\\
11-12 & HD 150193 & 11.7 & 1000 & 0.45 & 145(rot.) & Mag. & 1\\
 & " & 10.3 & 500 & 0.5 & 97 & MMT & 2\\
13-14 & HD 98922 & 11.7 & 1000 & 0.55 & -40(rot.) & Mag. & 1\\
15-16 & " & 10.3 & " & 0.55 & " & " & 1\\
17-18 & HD 104237 & 11.7 & 1000 & 0.55 & -60(rot.) & " & 1\\
19-20 & " & 10.3 & " & " & " & " & 1\\
20-21 & 51 Oph & 11.7 & 1000 & 0.11 & 135(rot.) & " & 1\\
22 & " & " & 500 & " & 45(rot.) & " & 1\\
23-24 & R CrA & 11.7 & 1000 & 0.11 & 0(rot.) & " & 1\\
 & " & " & " & " & 30(rot.) & " & 1\\
25-27 & AB Aur & 10.3 & 60 & 1.0 & 170 & MMTAO & 3\\
28-30 & " & " & " & " & 107 & " & 3\\
31-33 & " & " & " & " & 71 & " & 3\\
34-36 & " & " & " & " & 131 & " & 3\\
37-39 & " & " & " & " & 4 & " & 3\\
40-41 & V892 Tau & 11.7 & 1000 & 0.5 & 164 & MMT & 3\\
42-43 & " & " & " & " & 116 & " & 3\\
44-46 & " & 10.3 & 60 & 1.0 & 160 & MMTAO & 3\\
 & HD 100546 & 10.3 & 1000 & 0.5 & -80(rot.) & MMT & 4\\
 & " & " & " & " & -50(rot.) & " & 4\\
 & " & " & " & " & -24(rot.) & " & 4\\
 & " & " & " & " & 10(rot.) & " & 4\\
 & " & " & " & " & 40(rot.) & " & 4\\
 & " & 11.7 & 1000 & 0.5 & -80(rot.) & MMT & 4\\
 & " & " & " & " & -50(rot.) & " & 4\\
 & " & " & " & " & -24(rot.) & " & 4\\
 & " & " & " & " & 10(rot.) & " & 4\\
 & " & " & " & " & 40(rot.) & " & 4\\
 & " & 12.5 & 1000 & 0.5 & -80(rot.) & MMT & 4\\
 & " & " & " & " & -50(rot.) & " & 4\\
 & " & " & " & " & -24(rot.) & " & 4\\
 & " & " & " & " & 10(rot.) & " & 4\\
 & " & " & " & " & 40(rot.) & " & 4\\
 & HD 163296 & 10.3 & 500 & 0.5 & 94 & MMT & 2\\
 & " & " & " & " & 10 & " & 2\\
 & HD 179218 & 10.3 & 500 & 0.5 & 162 & MMT & 2\\
 & " & " & " & " & 87 & " & 2\\
 & " & 10.6 & 90 & 1.0 & 50 & MMTAO & 1\\
 & " & " & " & 1.0 & 60 & MMTAO & 1\\
 & " & " & 60 & 1.0 & 99 & MMTAO & 1\\
 & " & " & " & 1.0 & 103 & MMTAO & 1
\enddata
\tablecomments{References- 1) this paper; 2) \citet{hinz01}; 3) \citet{liu05};
4) \citet{liu03}\\
$^{*}$The PA probed by the nulling observation
depends on both the rotation of the interferometer baseline and the paralactic angle of the
object when observed.  For objects denoted 'rot.' the paralactic angle was not
recorded, so the rotation of the interferometer (degrees from an arbitrary
position) is noted so the reader can estimate the relative value of the 
PA for observations of each object.}
\end{deluxetable}

\clearpage
\begin{deluxetable}{ccccccc}
\rotate
\tablecaption{Instrumental and Source Nulls \label{tab-nulls}}
\tablewidth{0pt}
\tablehead{
\colhead{Star} &
\colhead{Instr. Null (\%)} &
\colhead{Cal. Null (\%)} &
\colhead{Source Null (\%)} &
\colhead{PA (or rotation)($\degr$)} &
\colhead{$\lambda$ ($\mu$m)} &
\colhead{Ref.}
}
\startdata
HR 5999 & $5.7 \pm 4.6$ & $5.6 \pm 2.5$ & $0.1 \pm 5.2$ & -165(rot.) & 11.7 & 1\\
" & $1.1 \pm 1.1$ & " & $-4.5 \pm 2.7$ & 110(rot.) & " & 1\\
KK Oph & $7.5 \pm 4.2$ & $5.6 \pm 2.5$ & $1.9 \pm 4.9$ & -150(rot.) & 11.7 & 1\\
" & $4.3 \pm 2.9$ & " & $-1.3 \pm 3.8$ & 120(rot.) & " & 1\\
DK Cha & $10.3 \pm 4.2$ & $8.4 \pm 1.5$ & $1.9 \pm 4.4$ & -170(rot) & 11.7 & 1\\
HD 150193 & $10.6 \pm 2.6$ & $6.9 \pm 0.2$ & $3.7 \pm 2.6$ & 145(rot.) & 11.7 & 1\\
" & $13 \pm 5$ & $13 \pm 2$ & $0 \pm 5$ & 97 & 10.3 & 2\\
HD 98922 & $5.9 \pm 3.3$ & $24.1 \pm 7.6$ & $-18.2 \pm 8.3$ & -40(rot.) & 11.7 & 1\\
" & $15.2 \pm 5.8$ & $12.3 \pm 3.9$ & $2.9 \pm 7.0$ & " & 10.3 & 1\\
HD 104237 & $9.2 \pm 6.7$ & $15.5 \pm 1.0$ & $-6.3 \pm 6.7$ & -60(rot.) & 11.7 & 1\\
" & $10.3 \pm 1.9$ & $15.4 \pm 3.2$ & $-5.1 \pm 3.7$ & " & 10.3 & 1\\
51 Oph & $11.9 \pm 1.1$ & $22.9 \pm 10.4$ & $-11.0 \pm 10.5$ & 1 & 11.7 & 1\\
" & $7.3 \pm 5.0$ & " & $-15.6 \pm 11.5$ & 2 & " & 1\\
R CrA & $19.8 \pm 5.1$ & $18.1 \pm 2.7$ & $1.7 \pm 5.8$ & 1 & 11.7 & 1\\
" & $26.4 \pm 1.9$ & " & $8.3 \pm 3.3$ & 2 & 11.7 & 1\\
AB Aur & $25.9 \pm 1.5$ & $5.1 \pm 0.2$ & $20.8 \pm 1.5$ & 170 & 10.3 & 3\\
" & $18.3 \pm 1.4$ & " & $13.2 \pm 1.4$ & 107 & " & 3\\
" & $24.1 \pm 2.9$ & " & $19.0 \pm 2.9$ & 71 & " & 3\\
" & $20.9 \pm 1.5$ & " & $15.8 \pm 1.5$ & 131 & " & 3\\
" & $31.7 \pm 3.1$ & " & $26.6 \pm 3.1$ & 4 & " & 3\\
V892 Tau & $24.2 \pm 2.2$ & $18.5 \pm 0.8$ & $5.7 \pm 2.3$ & 164 & 10.3 & 3\\
" & $37.9 \pm 7.2$ & $22.0 \pm 1.3$ & $15.9 \pm 7.3$ & 116 & 11.7 & 3\\
" & $36.4 \pm 0.2$ & " & $14.4 \pm 1.3$ & 160 & 11.7 & 3\\
HD 100546 & $45.5 \pm 1.6$ & $8.6 \pm 0.1$ & $36.9 \pm 1.6$ & -80(rot) & 10.3 & 4\\
" & $39.9 \pm 7.0$ & " & $31.3 \pm 7.0$ & -50(rot) & " & 4\\
" & $41.9 \pm 2.8$ & " & $33.3 \pm 2.8$ & -24(rot) & " & 4\\
" & $29.8 \pm 8.9$ & " & $21.2 \pm 8.9$ & 10(rot) & " & 4\\
" & $31.6 \pm 0.6$ & $12.5 \pm 1.8$ & $19.1 \pm 1.9$ & 40(rot) & " & 4\\
" & $40.0 \pm 2.3$ & $6.4 \pm 2.4$ & $33.6 \pm 3.3$ & -80(rot) & 11.7 & 4\\
" & $26.1 \pm 2.7$ & " & $19.7 \pm 3.6$ & -50(rot) & " & 4\\
" & $37.4 \pm 2.5$ & $8.6 \pm 2.0$ & $28.8 \pm 3.2$ & -24(rot) & " & 4\\
" & $31.8 \pm 2.6$ & " & $23.2 \pm 3.3$ & 10(rot) & " & 4\\
" & $24.3 \pm 2.0$ & $9.8 \pm 2.2$ & $14.5 \pm 3.0$ & 40(rot) & " & 4\\
" & $35.5 \pm 2.8$ & $5.6 \pm 1.8$ & $29.9 \pm 3.3$ & -80(rot) & 12.5 & 4\\
" & $48.8 \pm 1.3$ & " & $43.2 \pm 2.2$ & -50(rot) & " & 4\\
" & $30.9 \pm 1.7$ & $6.1 \pm 0.1$ & $24.8 \pm 1.7$ & -24(rot) & " & 4\\
" & $26.4 \pm 1.6$ & " & $20.3 \pm 1.6$ & 10(rot) & " & 4\\
" & $16.3 \pm 5.9$ & $10.3 \pm 2.1$ & $6.0 \pm 6.3$ & 40(rot) & " & 4\\
HD 163296 & $12 \pm 7$ & $13 \pm 2$ & $-1 \pm 7$ & 94 & 10.3 & 2\\
" & $16 \pm 2$ & " & $3 \pm 3$ & 10 & " & 2\\
HD 179218 & $16 \pm 2$ & $13 \pm 2$ & $3 \pm 3$ & 162 & 10.3 & 2\\
" & $14 \pm 2$ & " & $1 \pm 3$ & 87 & " & 2\\
" & $10.5 \pm 2.4$ & $3.7 \pm 0.6$ & $6.8 \pm 2.5$ & 50 & 10.6 & 1\\
" & $8.5 \pm 1.8$ & $3.6 \pm 0.8$ & $4.9 \pm 2.0$ & 60 & " & 1\\
" & $7.2 \pm 1.2$ & " & $3.6 \pm 1.4$ & 99 & " & 1\\
" & $11.4 \pm 2.5$ & $5.9 \pm 1.2$ & $5.5 \pm 2.8$ & 103 & " & 1
\enddata
\tablecomments{References- 1) this paper; 2) \citet{hinz01}; 3) \citet{liu05};
4) \citet{liu03}}
\end{deluxetable}

\clearpage
\begin{deluxetable}{cccccc}
\rotate
\tablecaption{Best Fit Models\label{tab-models}}
\tablewidth{0pt}
\tablehead{
\colhead{Name} &
\colhead{$\lambda$ ($\mu$m)} &
\colhead{Gaussian FWHM (AU)} &
\colhead{Ring diameter (AU)} &
\colhead{Flaring Exp.$^{*}$} &
\colhead{Ref.}
}
\startdata
HR 5999 & 11.7 & $<19$ & $<26$ & $<0.06$ & 1\\
KK Oph & 11.7 & $<14$ & $<19$ & $<0.06$ & 1\\
DK Cha & 11.7 & $<17$ & $<23$ & $<0.05$ & 1\\
HD 150193 & 11.7 & $<12$ & $<16$ & $<0.01$ & 2\\
HD 98922 & 11.7 & $<48$ & $<68$ & $<0.04$ & 1\\
HD 104237 & 11.7 & $<12$& $<16$ & $<0.07$ & 1\\
51 Oph & 11.7 & $<17$ & $<23$ & $<0.08$ & 1\\
R CrA & 11.7 & $15 \pm 4$ & $20 \pm 4$ & 0.02-0.07 & 1\\
AB Aur & 10.3 & $27 \pm 3$ & $30 \pm 3$ & 0.1-0.15 & 3\\
V892 Tau & 10.3 & $14 \pm 2$ & $20 \pm 3$ & 0.05-0.15 & 3\\
" & 11.7 & $23 \pm 5$ & $31 \pm 6$ & 0.05-0.15 & 3\\
HD 100546 & 10.3 & $24 \pm 3$ & $26 \pm 3$ & 0.08-0.18 & 4\\
" & 11.7 & $25 \pm 3$ & $27 \pm 3$ & 0.07-0.18 & 4\\
" & 12.5 & $30 \pm 3$ & $33 \pm 3$ & 0.02-0.2 & 4\\
HD 163296 & 11.7 & $<11$ & $<15$ & $<0.08$ & 2\\
HD 179218 & 10.6 & $20 \pm 4$ & $27 \pm 5$ & 0.01-0.05 & 1
\enddata
\tablecomments{References- 1) this paper; 2) \citet{hinz01}; 3) \citet{liu05};
4) \citet{liu03} \\
$^{*}$ in the range 0 to 2/7; see \S \ref{sec-disc2}}
\end{deluxetable}

\clearpage

\clearpage

\clearpage


\begin{thebibliography}

\bibitem[Bouwman et al.(2003)]{bouwman}
Bouwman, J., de Koter, A., Dominik, C., \& Waters, L.~B.~F.~M.\ 2003, \aap,
401, 577

\bibitem[Bracewell(1978)]{bracewell} Bracewell, R.~N.\ 1978,
\nat, 274, 780

\bibitem[Brusa et al.(2003)]{brusa} Brusa, G., et al.\ 2003, 
\procspie, 4839, 691

\bibitem[Calvet et al.(1991)]{calvet} Calvet, N., Patino, A., 
Magris, G.~C., \& D'Alessio, P.\ 1991, \apj, 380, 617 

\bibitem[Chen \& Jura(2003)]{cj03} Chen, C.~H.~\& Jura, M.\
2003, \apj, 591, 267

\bibitem[Chiang \& Goldreich(1999)]{cg99} Chiang, E.~I., \& 
Goldreich, P.\ 1999, \apj, 519, 279 

\bibitem[Chiang \& Goldreich(1997)]{cg97} Chiang, E.~I., \&
Goldreich, P.\ 1997, \apj, 490, 368

\bibitem[Cox(2000)]{cox} Cox, A.~N.\ 2000, Allen's 
astrophysical quantities, 4th ed.~Publisher: New York: AIP Press; Springer, 
2000.~Editedy by Arthur N.~Cox.~ ISBN: 0387987460

\bibitem[de Zeeuw et al.(1999)]{dezeeuw} de Zeeuw, P.~T., 
Hoogerwerf, R., de Bruijne, J.~H.~J., Brown, A.~G.~A., \& Blaauw, A.\ 1999, 
\aj, 117, 354 

\bibitem[Dullemond(2002)]{d02} Dullemond, C.~P.\ 2002, 
\aap, 395, 853 

\bibitem[Dullemond et al.(2001)]{ddn01} Dullemond, C.~P., 
Dominik, C., \& Natta, A.\ 2001, \apj, 560, 957

\bibitem[Eisner et al.(2003)]{eis03} Eisner, J.~A., Lane,
B.~F., Akeson, R.~L., Hillenbrand, L.~A., \& Sargent, A.~I.\ 2003, \apj,
588, 360

\bibitem[Fukagawa et al.(2004)]{f04} Fukagawa, M., et al.\
2004, \apjl, 605, L53

\bibitem[Hamaguchi et al.(2005)]{hama} Hamaguchi, K., 
Yamauchi, S., \& Koyama, K.\ 2005, \apj, 618, 360

\bibitem[Hartmann, Kenyon, \& Calvet(1993)]{hkc93} Hartmann,
L., Kenyon, S.~J., \& Calvet, N.\ 1993, \apj, 407, 219

\bibitem[Hillenbrand et al.(1992)]{hill92}
Hillenbrand, L.~A., Strom, S.~E., Vrba, F.~J., \& Keene, J.\ 1992, \apj,
397, 613

\bibitem[Hinz (2001)]{hinz_thesis} Hinz, P.~M.\ 2001, Ph.D. Thesis,
University of Arizona

\bibitem[Hinz et al.(2001)]{hinz01} Hinz, P.~M., Hoffmann,
W.~F., \& Hora, J.~L.\ 2001, \apjl, 561, L131

\bibitem[Hoffmann et al. (1998)]{hoffmann} Hoffmann, W.~F., Hora,
J.~L., Fazio, G.~G., Deutsch, L.~K., \& Dayal, A.\ 1998, \procspie, 3354,
647

\bibitem[Kenyon \& Hartmann(1987)]{kh87} Kenyon, S.~J.~\&
Hartmann, L.\ 1987, \apj, 323, 714

\bibitem[Lada \& Adams(1992)]{la92} Lada, C.~J.~\& Adams,
F.~C.\ 1992, \apj, 393, 278

\bibitem[Leinert et al.(2004)]{leinert} Leinert, C., et al.\ 
2004, \aap, 423, 537

\bibitem[Liu et al.(2005)]{liu05} Liu, W.~M., Hinz, P.~M.,
Hoffmann, W.~F., Brusa, G., Miller, D., \& Kenworthy, M.~A.\ 2005, \apjl,
618, L133

\bibitem[Liu et al.(2003)]{liu03} Liu, W.~M., Hinz, P.~M.,
Meyer, M.~R., Mamajek, E.~E., Hoffmann, W.~F., \& Hora, J.~L.\ 2003, \apjl,
598, L111

\bibitem[Malfait et al.(1998)]{malfait} Malfait, K., Waelkens, 
C., Waters, L.~B.~F.~M., Vandenbussche, B., Huygen, E., \& de Graauw, 
M.~S.\ 1998, \aap, 332, L25

\bibitem[Mamajek et al.(2002)]{mamajek} Mamajek, E.~E., Meyer, 
M.~R., \& Liebert, J.\ 2002, \aj, 124, 1670 

\bibitem[Mannings \& Sargent(1997)]{ms97} Mannings, V.~\&
Sargent, A.~I.\ 1997, \apj, 490, 792

\bibitem[Marinas et al. (2006)]{marinas} Marinas, N., et al.\ 2006, accepted
ApJ, astro-ph \#0609119

\bibitem[Marsh et al.(1995)]{marsh95} Marsh, K.~A., Van Cleve,
J.~E., Mahoney, M.~J., Hayward, T.~L., \& Houck, J.~R.\ 1995, \apj, 451,
777

\bibitem[Meeus et al.(2001)]{meeus} Meeus, G., Waters,
L.~B.~F.~M., Bouwman, J., van den Ancker, M.~E., Waelkens, C., \& Malfait,
K.\ 2001, \aap, 365, 476

\bibitem[Millan-Gabet et al.(2001)]{m-g06} Millan-Gabet R., Malbet, F., Akeson, R., Leinert, C., Monnier, J., \& Waters, R.\ 2006, Protostars and Planets V.

\bibitem[Millan-Gabet et al.(2001)]{m-g01}
Millan-Gabet, R., Schloerb, F.~P., \& Traub, W.~A.\ 2001, \apj, 546, 358

\bibitem[Miroshnichenko et al.(1999)]{m99} Miroshnichenko, A.,
Ivezi{\' c} , {\v Z}., Vinkovi{\' c} , D., \& Elitzur, M.\ 1999,
\apjl, 520, L115

\bibitem[Natta et al.(2000)]{natta} Natta, A., Grinin, V., \& 
Mannings, V.\ 2000, Protostars and Planets IV, 559

\bibitem[Perryman et al.(1997)]{perryman} Perryman, M.~A.~C., et 
al.\ 1997, \aap, 323, L49

\bibitem[Press et al.(1992)]{num_rec} Press, W.~H., Teukolsky, 
S.~A., Vetterling, W.~T., \& Flannery, B.~P.\ 1992, Cambridge: University 
Press, |c1992, 2nd ed.

\bibitem[The et al.(1994)]{the94} The, P.~S., de Winter, D.,
\& Perez, M.~R.\ 1994, \aaps, 104, 315

\bibitem[van den Ancker et al.(1998)]{vdA} van den Ancker, 
M.~E., de Winter, D., \& Tjin A Djie, H.~R.~E.\ 1998, \aap, 330, 145 

\bibitem[Vinkovi{\'c} et al.(2006)]{vink06} Vinkovi{\'c}, D., 
Ivezi{\'c}, {\v Z}., Jurki{\'c}, T., \& Elitzur, M.\ 2006, \apj, 636, 348 

\bibitem[Vinkovi{\'c} et al.(2003)]{vink03} Vinkovi{\'c}, D., 
Ivezi{\'c}, {\v Z}., Miroshnichenko, A.~S., \& Elitzur, M.\ 2003, \mnras, 
346, 1151 

\bibitem[Waelkens et al.(1996)]{wael} Waelkens, C., et al.\ 
1996, \aap, 315, L245

\end{thebibliography}
\end{document}